\documentclass[num-refs]{wiley-article}
\usepackage[utf8]{inputenc}
\usepackage{lipsum}
\usepackage{booktabs}
\PassOptionsToPackage{hyphens}{url}
\usepackage{hyperref}
\usepackage{multirow}
\usepackage{underscore}
\usepackage{xcolor}
\usepackage{pgfplots}
\pgfplotsset{compat=1.17}
\usepackage{subcaption}
\usepackage{adjustbox}
\usepackage{microtype}
\usepackage{tikz}
\usepackage{xcolor}
\definecolor{invalid}{RGB}{255,0,0}
\definecolor{valid}{RGB}{0,128,0}
\definecolor{padding}{RGB}{128,128,128}
\definecolor{mask}{RGB}{0,0,255}
\usepackage[per-mode=symbol]{siunitx}
\usepackage{tikzsymbols}
\usepackage{float}
\usepackage{afterpage}
\newcommand{\repchar}{\fbox{?}}
\usepackage{tikz}
\usetikzlibrary{shapes,arrows,positioning}

\usepackage{listings}
\lstdefinestyle{customcpp}{%
  belowcaptionskip=1\baselineskip,
  breaklines=true,
  xleftmargin=\parindent,
  language=C++,
  showstringspaces=false,
  basicstyle=\linespread{0.4}\footnotesize\ttfamily,
  keywordstyle=\bfseries\color{green!40!black},
  numberstyle=\tiny,
  commentstyle=\itshape\color{purple!40!black},
  identifierstyle=\bfseries\color{black},
  stringstyle=\color{red},
  emph={int,char,double,float,unsigned},
  emphstyle=\color{blue},
  morekeywords={uint64_t,uint32_t,__m256i,__m128i,UINT64_C,uint16x8_t,vdupq_n_u16,vld1q_u16,vcleq_u16,vaddq_u16,vextq_u16,vbicq_u16,vorrq_u16,vbslq_u16,vst1q_u16}
}

\lstdefinestyle{customasmarm}{%
  belowcaptionskip=1\baselineskip,
  breaklines=true,
  xleftmargin=\parindent,
  language=[ARM]{Assembler},
  showstringspaces=false,
  basicstyle=\linespread{0.4}\small\ttfamily,
  keywordstyle=\bfseries\color{green!40!black},
  numberstyle=\tiny,
  commentstyle=\itshape\color{purple!40!black},
  identifierstyle=\bfseries\color{black},
  stringstyle=\color{red},
}

\definecolor{bblue}{HTML}{4F81BD}
\definecolor{rred}{HTML}{C0504D}
\definecolor{ggreen}{HTML}{9BBB59}
\definecolor{ppurple}{HTML}{9F4C7C}
\definecolor{light-gray}{gray}{0.95}

\hypersetup{
    colorlinks,
    linkcolor={red!50!black},
    citecolor={blue!50!black},
    urlcolor={black}
}

\newcommand{\codepoint}[1]{\texttt{U+\MakeUppercase{#1}}}
\newcommand{\codepointrange}[2]{\texttt{U+\MakeUppercase{#1}}--\texttt{U+\MakeUppercase{#2}}}
\newcommand*{\restartrowcolors}{%
  \ifhmode\unskip\fi
  \vadjust{%
    \global\rownum=0 %
  }%
}

\title{Fixing ill-formed UTF-16 strings with SIMD instructions}
\author[1]{Robert Clausecker}
\author[2]{Daniel Lemire}
\affil[1]{Zuse Institute Berlin, Germany}
\affil[2]{Université du Québec (TELUQ), Montreal, Quebec, H2S 3L5, Canada}
\corraddress{Daniel Lemire, Université du Québec (TELUQ), Montreal, Quebec, H2S 3L5, Canada}
\corremail{daniel.lemire@teluq.ca}
\fundinginfo{Natural Sciences and Engineering Research Council of Canada, Grant Number: RGPIN-2024-03787}
\runningauthor{Robert Clausecker and Daniel Lemire}

\begin{document}
\maketitle

\begin{abstract}
UTF-16 is a widely used Unicode encoding representing characters with one or two 16-bit code units. The format  relies on surrogate pairs to encode characters beyond the Basic Multilingual Plane, requiring a high surrogate followed by a low surrogate. Ill-formed UTF-16 strings---where surrogates are mismatched---can arise from data corruption or improper encoding, posing security and reliability risks. Consequently, programming languages such as JavaScript include functions to fix ill-formed UTF-16 strings by replacing mismatched surrogates with the Unicode replacement character (U+FFFD). 
We propose using Single Instruction, Multiple Data (SIMD) instructions to handle multiple code units in parallel, enabling faster and more efficient execution. Our software is part of the Google JavaScript engine (V8) and thus part of several major Web browsers.
\end{abstract}

\section{Introduction}
Unicode is the standard for text representation in modern software, supporting over one million 
characters across diverse writing systems.
Unicode assigns each character a unique code point from \codepoint{0000} to \codepoint{10FFFF}, 
organized into 17~planes. The Basic Multilingual Plane (BMP, \codepointrange{0000}{FFFF}) contains the 
most commonly used characters, while supplementary planes encode less frequent characters, such as 
ideographs or emojis.
\emph{Unicode Transformation Formats} (UTF) encode these characters in binary form.
One of the primary Unicode Transformation Formats, UTF-16, is used by platforms such as Microsoft Windows, Java, and JavaScript for internal string representation. 
UTF-16 encodes characters in the Basic Multilingual Plane (BMP, \codepointrange{0000}{FFFF}) using a single 16-bit code unit, while characters in supplementary planes (\codepointrange{10000}{10FFFF}) are encoded as surrogate pairs: a high surrogate (\codepointrange{D800}{DBFF}) followed by a low surrogate (\codepointrange{DC00}{DFFF}).
The code point is computed from a surrogate pair \((h, l)\) as:
\[
\text{Code point} = ((h - \mathrm{0xD800}) \ll 10) + (l - \mathrm{0xDC00}) + \mathrm{0x10000}
\]
where \(h\) is the high surrogate and \(l\) is the low surrogate.

Ill-formed UTF-16 strings occur when surrogate pairs are mismatched, such as a high surrogate not followed by a low surrogate or a low surrogate appearing without a preceding high surrogate. Such errors can result from data corruption, improper encoding conversions, or malicious inputs, potentially leading to security vulnerabilities or application crashes~\cite{utr36,cve20122135}. For example, the UTF-16 decoder of CPython~3.1 to~3.3 failed to update an internal bound after invoking its error handler---which is triggered by ill-formed sequences such as an unpaired surrogate---so that ill-formed input caused the decoder to read past the end of its buffer, disclosing process memory or crashing the application~\cite{cve20122135}. To mitigate these issues, mismatched surrogates are typically replaced with the Unicode replacement character (\codepoint{FFFD}), ensuring robust text processing and signaling errors to developers or users.

Conventional scalar algorithms for fixing ill-formed UTF-16 strings process code units sequentially, as shown in Fig.~\ref{fig:scalar}. While straightforward, these methods are computationally expensive for large strings, limiting their suitability for high-throughput applications like web browsers or databases. Modern processors, including ARM and x64 architectures, support Single Instruction, Multiple Data (SIMD) instructions, enabling parallel processing of multiple data elements. We propose a SIMD-based algorithm to accelerate UTF-16 correction, leveraging ARM NEON and x64 Streaming SIMD Extensions (SSE) instructions to process 16-bit code units in blocks, achieving significant performance gains.

\begin{figure}
    \centering
    \begin{tabular}{l}
    \lstinputlisting[style=customcpp]{code/scalar\_utf16.c}
    \end{tabular}
    \caption{Scalar C function to replace invalid UTF-16 surrogates with the replacement character}
    \label{fig:scalar}
\end{figure}

\begin{figure}
    \centering
    \includegraphics[width=0.7\linewidth]{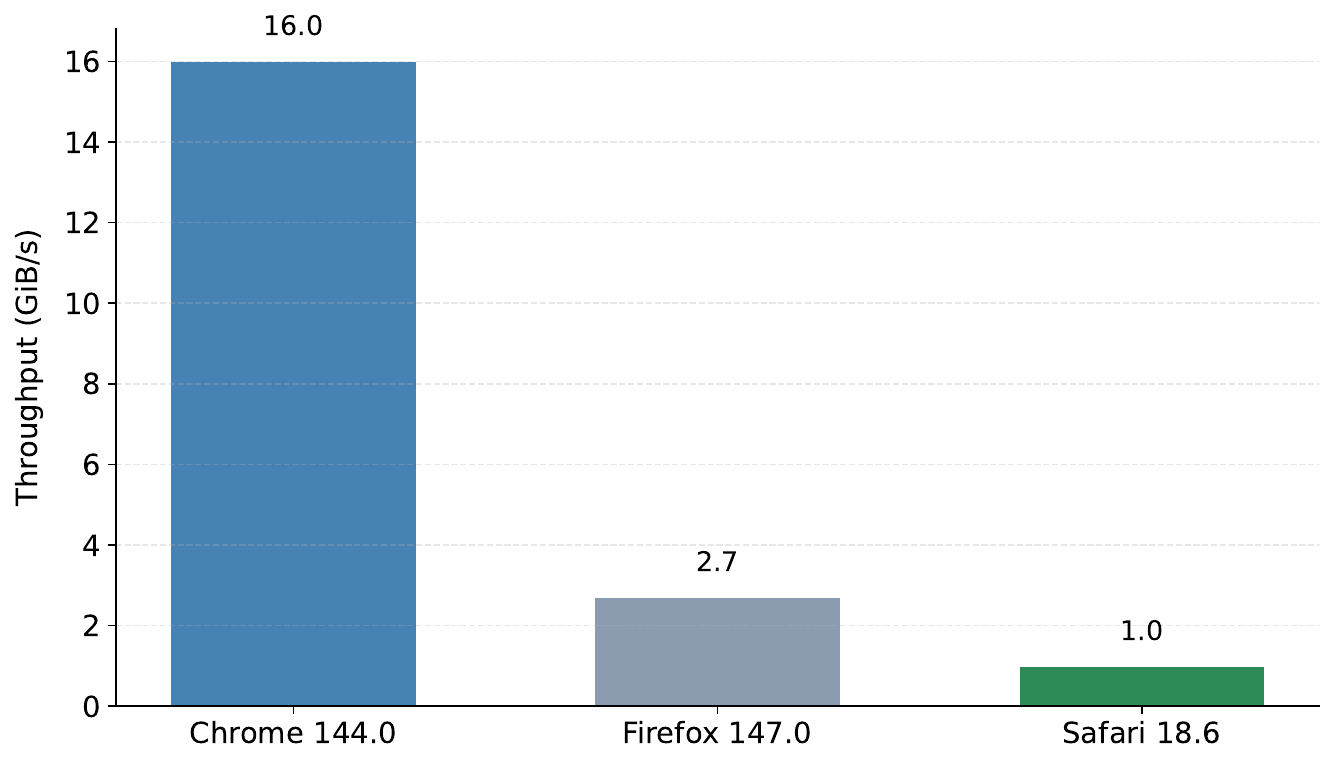}
    \caption{Speed of the \texttt{String.toWellFormed} JavaScript function using a 1024-character string on different browsers with an Apple~M4 processor.
    The string is random, made of 5\% low surrogates, 5\% high surrogates, the rest being ASCII characters.}
    \label{fig:browser}
\end{figure}

Our contributions include a novel SIMD algorithm for in-place and copy-based UTF-16 correction, implementations optimized for ARM NEON and x64 SSE, and a comprehensive experimental evaluation demonstrating up to eight-fold speedups over scalar methods. 
Our software is part of the open-source simdutf library, a widely used library: it helps ensure reproducibility and practical applicability.
Our method has been deployed in Chromium browsers (Chrome, Edge, Brave and others): Fig.~\ref{fig:browser} illustrates the superior performance (\SI{16}{\gibi\byte\per\second}, several times faster than competing browsers.\footnote{We make available an online benchmark at \url{https://lemire.github.io/browserwellformed/}.}

\section{Related work}
Unicode processing has received attention for tasks like validation and transcoding, but correcting ill-formed UTF-16 strings is less explored. Keiser and Lemire~\cite{keiser2020validating} developed SIMD-based UTF-8 validation algorithms, processing multiple bytes in parallel. Lemire and Mu{\l}a~\cite{lemire2022transcoding} proposed SIMD-accelerated UTF-8 to UTF-16 transcoding, achieving gigabytes-per-second throughput using ARM NEON and x64 SSE instructions.  More recently, Schröder et al.~\cite{schroder2024validating} proposed a SIMD-based algorithm for validating CESU-8 encoded text, an encoding scheme combining UTF-8’s ASCII compatibility with UTF-16’s binary order. Their work, utilizing x86, ARM, and PowerPC SIMD instructions, achieves a sevenfold performance improvement over conventional validation methods, as demonstrated on datasets with ASCII, Hangul, and random text. These works focus on validation or format conversion, not in-place correction of UTF-16 strings.

Cameron~\cite{cameron2008case} introduced bit-stream-based SIMD processing for UTF-8 to UTF-16 transcoding, but without addressing surrogate pair correction.
Similarly, Lemire and Muła~\cite{lemire2022transcoding} proposed SIMD-accelerated transcoding from UTF-8 to UTF-16, achieving gigabyte-per-second throughput using ARM NEON and x64 SSE instructions.
Clausecker and Lemire~\cite{clausecker2023transcoding} further advanced transcoding performance by leveraging Advanced Vector Extensions~512 (AVX-512) instructions to process 512-bit registers, achieving over \SI{5}{\gibi\byte\per\second} for UTF-8 to UTF-16 transcoding of Chinese text with fewer than two CPU instructions per character.
However, these algorithms only do transcoding and not in-place correction.

Scalar UTF-16 validation is part of standard libraries (e.g., ICU, Boost), but these implementations process code units sequentially, lacking SIMD optimization. Our work builds on SIMD techniques from prior studies, adapting them to the specific problem of UTF-16 surrogate correction, and introduces novel optimizations for ARM NEON.

More broadly, SIMD instructions are  used to accelerate many string operations.  For exact string matching, a naive approach leveraging these instructions can be remarkably effective. As demonstrated by Tarhio et al., an algorithm that compares characters in a special order and utilizes SIMD instructions can outperform more complex conventional algorithms~\cite{tarhio2017technology}. Similarly, Fiori et al.~\cite{fiori2022approximate} and Chhabra et al.~\cite{chhabra2025string} introduce novel algorithms for the approximate matching problem between strings. We can also use SIMD instructions to parse strings more quickly, such as JSON strings~\cite{keiser2023demand,langdale2019parsing}, XML~\cite{10.1145/1463788.1463811}, DNS records~\cite{simdzone}, and so forth.

\section{SIMD algorithm}

SIMD is a parallel processing technique that allows a single instruction to operate on multiple data elements simultaneously. The core motivation behind SIMD is to exploit data-level parallelism, enabling processors to perform the same operation—such as addition or comparison—on multiple data points in a single cycle, thereby reducing execution time. By processing multiple elements in parallel, SIMD not only accelerates computation but also improves power efficiency~\cite{7503679,wassenberg2022vectorized}, as it reduces the number of instruction cycles needed. 

Our algorithm processes UTF-16 strings in blocks of at least 8~code units (16~bytes) using SIMD instructions, correcting mismatched surrogates by replacing them with \codepoint{FFFD}. It supports both in-place correction (input equals output) and copy-based correction (input and output are distinct buffers). The algorithm checks for valid surrogate pairs by examining high and low surrogates across block boundaries, using a lookback mechanism to handle pairs spanning blocks.

\begin{figure}
    \centering
    \begin{tabular}{l}
    \lstinputlisting[style=customcpp]{code/simd\_utf16.c}
    \end{tabular}
    \caption{Generic SIMD function to replace invalid UTF-16 surrogates with the replacement character}
    \label{fig:genericsimd}
\end{figure}

\subsection{Architecture-independent algorithm}
Our algorithm (Fig.~\ref{fig:genericsimd}) works in three steps: first, we load two vectors of 16-bit code units with an offset of one word from the input.  The earlier vector (the \emph{lookback}) is checked for the presence of high surrogates, the later vector (the \emph{block}) for the presence of low surrogates, yielding two vectors of booleans \textit{lb\_is_high} and \textit{block\_is\_low}.  In the second step, we compute the element-wise exclusive-or of these two vectors of booleans.  As the two vectors have an offset of one word, this checks if a high surrogate is not followed by a low surrogate or vice versa.  If the result is all zeros, the current block is correctly sequenced and we copy it to the output.\footnote{When the processing is done in-place, the copy is omitted.}

Otherwise we proceed to the third step, which involves detecting which specific 16-bit code units are illegally sequenced and replacing them with the Unicode replacement character~\codepoint{FFFD} prior to writing the block vector to the output buffer.  We compute which high surrogates are not followed by a low surrogate by taking the element-wise and of \textit{lb\_is_high} with the complement of \textit{block\_is\_low} and shifting the resulting vector to the right by one element (of two bytes) to have the result correspond to elements in \emph{block} instead of \emph{lookback}.  Likewise, low surrogates not preceded by high surrogates are found by taking the element-wise and of the complement of \textit{lb\_is_high} with \textit{block\_is\_low}.  The element-wise or of these two vectors is taken to get \emph{block\_illseq}, indicating illegally sequenced code units in \emph{block}, which are replaced with~\codepoint{FFFD} using a blend operation prior to writing them back.


We perform this algorithm on each vector-sized block of input in turn.  As it is idempotent, a tail of less than the block size is handled by performing a final iteration aligned to the end of the input, possibly overlapping the penultimate iteration.  Two cases (input starts with low surrogate or ends with high surrogate) are not caught by the vectorised procedure and must be taken care of manually.  Likewise, input that is shorter than one vector and one element cannot be processed by our algorithm and must fall back to the scalar procedure from Fig.~\ref{fig:scalar}.


\begin{figure}
\centering
\begin{tikzpicture}
  \newcommand{\fixedHeight}{0.33cm}
  \tikzstyle{cell}=[rectangle, draw, minimum width=1.35cm, minimum height=\fixedHeight,  font=\ttfamily\small]
  \tikzstyle{invalid}=[cell, fill=red!20]
  \tikzstyle{valid}=[cell, fill=green!20]
  \tikzstyle{flag}=[cell, fill=blue!5]
  \tikzstyle{label}=[font=\bfseries\small, rotate=45, anchor=west]
  \tikzstyle{varlabel}=[font=\ttfamily\small, rotate=45, anchor=west]

  \node[label]    (h0) at (-0.3, 0.55) {Index};
  \node[label]    (h1) at (1.1, 0.55) {Input};
  \node[varlabel] (h2) at (2.5, 0.55) {lookback};
  \node[varlabel] (h3) at (3.9, 0.55) {lb\_masked};
  \node[varlabel] (h4) at (5.3, 0.55) {block\_masked};
  \node[varlabel] (h5) at (6.7, 0.55) {lb\_is\_high};
  \node[varlabel] (h6) at (8.1, 0.55) {block\_is\_low};
  \node[varlabel] (h7) at (9.5, 0.55) {illseq};
  \node[label]    (h8) at (10.9, 0.55) {Output};
  \newcommand{\offset}{20.8}

  \foreach \i/\inVal/\lbVal/\lbmVal/\bmVal/\lbhVal/\bilVal/\illVal/\outVal/\sty [count=\y] in {
    0/0048/0000/0000/0000/0/0/0/0048/cell,
    1/0065/0048/0000/0000/0/0/0/0065/cell,
    2/006C/0065/0000/0000/0/0/0/006C/cell,
    3/006C/006C/0000/0000/0/0/0/006C/cell,
    4/006F/006C/0000/0000/0/0/0/006F/cell,
    5/002C/006F/0000/0000/0/0/0/002C/cell,
    6/0020/002C/0000/0000/0/0/0/0020/cell,
    7/0077/0020/0000/0000/0/0/0/0077/cell,
    8/006F/0077/0000/0000/0/0/0/006F/cell,
    9/0072/006F/0000/0000/0/0/0/0072/cell,
    10/D800/0072/0000/D800/0/0/0/FFFD/invalid,
    11/006C/D800/D800/0000/1/0/1/006C/cell,
    12/0064/006C/0000/0000/0/0/0/0064/cell,
    13/0021/0064/0000/0000/0/0/0/0021/cell,
    14/0048/0021/0000/0000/0/0/0/0048/cell,
    15/0065/0048/0000/0000/0/0/0/0065/cell,
    16/006C/0065/0000/0000/0/0/0/006C/cell,
    17/006C/006C/0000/0000/0/0/0/006C/cell,
    18/006F/006C/0000/0000/0/0/0/006F/cell,
    19/0020/006F/0000/0000/0/0/0/0020/cell,
    20/0077/0020/0000/0000/0/0/0/0077/cell,
    21/DC00/0077/0000/DC00/0/1/1/FFFD/invalid,
    22/006F/DC00/DC00/0000/0/0/0/006F/cell,
    23/0072/006F/0000/0000/0/0/0/0072/cell,
    24/006C/0072/0000/0000/0/0/0/006C/cell,
    25/0064/006C/0000/0000/0/0/0/0064/cell,
    26/0021/0064/0000/0000/0/0/0/0021/cell,
    27/0048/0021/0000/0000/0/0/0/0048/cell,
    28/0065/0048/0000/0000/0/0/0/0065/cell,
    29/D83D/0065/0000/D800/0/0/0/D83D/valid,
    30/DE0A/D83D/D800/DC00/1/1/0/DE0A/valid,
    31/0000/DE0A/DC00/0000/0/0/0/0000/cell
  } {
    \node[font=\ttfamily\small, anchor=east] at (0.2, -\y*\fixedHeight + \offset) {\i};
    \node[\sty]  at (1.4,  -\y*\fixedHeight + \offset) {\inVal};
    \node[cell]  at (2.8,  -\y*\fixedHeight + \offset) {\lbVal};
    \node[cell]  at (4.2,  -\y*\fixedHeight + \offset) {\lbmVal};
    \node[cell]  at (5.6,  -\y*\fixedHeight + \offset) {\bmVal};
    \node[flag]  at (7.0,  -\y*\fixedHeight + \offset) {\lbhVal};
    \node[flag]  at (8.4,  -\y*\fixedHeight + \offset) {\bilVal};
    \node[flag]  at (9.8,  -\y*\fixedHeight + \offset) {\illVal};
    \node[\sty]  at (11.2, -\y*\fixedHeight + \offset) {\outVal};
  }
\end{tikzpicture}
\caption{Values of the variables of \texttt{utf16fix\_block} (Fig.~\ref{fig:genericsimd}) while
processing the example string with a vector length of 64~bytes.
The mismatched surrogates at indices 10 and 21 are marked in red, the correctly sequenced
surrogate pair at indices 29--30 in green.}
    \label{fig:examplevars}
\end{figure}

\paragraph{Example}

We shall demonstrate the operation of \texttt{utf16fix\_block} function of Fig.~\ref{fig:genericsimd} with a vector length of 64~bytes on the input string
\begin{equation*}
\texttt{Hello, wor\repchar{}ld!Hello w\repchar{}orld!He\Smiley[][yellow]}
\end{equation*}
where \texttt{\repchar{}} is a mismatched surrogate.
The UTF-16 string has a length of 31~code units and is surrounded by \codepoint{0000} on both sides.
The values of the variables of \texttt{utf16fix\_block} as it processes our string are shown in Fig.~\ref{fig:examplevars}.
The input, shown in the first column, contains mismatched surrogates \texttt{D800} at position~10 and \texttt{DC00} at position~21 marked in red.
There is also a correctly sequenced surrogate pair \texttt{D83D DE0A} at positions 29--30 encoding 
\codepoint{1F60A} \textsc{smiling face with smiling eyes} marked in green.

The \emph{lookback} vector with its leading \codepoint{0000} is shown in the second column.
The third and fourth columns show \emph{lookback} and \emph{block} masked with~\texttt{FC00}, from which are computed the \emph{lb\_is\_high} and \emph{block\_is\_low} masks shown in the fifth and sixth columns, flagging high surrogates in \emph{lookback} and low surrogates in \emph{block}.
The exclusive-or of these masks forms the \emph{illseq} vector, which indicates the presence, but not the precise location of mismatched surrogates.
In this case, the surrogate~\texttt{D800} at position~10 is incorrectly indicated at position 11.
Correct location information is only determined if \emph{illseq} is found to be not all zeros.
The mismatched surrogates are then replaced with \codepoint{FFFD} \textsc{replacement character}, 
leaving the correctly sequenced surrogates at position~29--30 alone.

\paragraph{Discussion}
Three subtle design decisions influence the  procedure.
The first decision is to track high and low surrogates in two overlapping vectors of input instead of using one vector and shifting the resulting masks \emph{lb\_is\_high} and \emph{block\_is\_low} to find mismatches.
The second decision is to perform two loads from the input buffer to obtain these overlapping vectors \emph{lookback} and \emph{block} instead of performing only one load per iteration and slicing out \emph{lookback} from the previous and the current \emph{block} vectors.
The third decision is to make the algorithm branchy,\footnote{A branching algorithm executes different code paths based on the input data for a given input size, while a branchless algorithm consistently follows the same code path regardless of the data content.} skipping the correction of illegally sequenced surrogates unless those actually appear.

We define the operational intensity of our main routine as the ratio of the number of arithmetic-logic SIMD instructions 
(e.g., comparisons, bitwise-AND) to the number of SIMD register loads and stores. In the best scenario, we need at least five~arithmetic-logic SIMD instructions for each iteration compared to at most two load operations and up to one store operation. 

Using overlapping vectors instead of shifting masks avoids having to carry over values from the previous iteration for ``lookback'' or ``lookahead.''
Loading twice instead of assembling \emph{lookback} from two \emph{block} vectors removes the need for shuffle operations at the cost of an extra load per iteration. 
On most processors, this is a good tradeoff. Indeed, current processors are often capable of retiring two or more SIMD load instructions per cycle~\cite{agner4}. Meanwhile, in the best scenario, 
we need at least five~arithmetic-logic SIMD instructions (e.g., comparisons or bitwise-AND operations) for each iteration. Thus we have a relatively high \emph{operational intensity}~\cite{10.1145/1498765.1498785} even with two load operations per iteration: we have many more arithmetic-logic instructions than memory operations. Increasing the operational intensity further would diminish the performance.

A potential downside of loading SIMD registers with an offset of two bytes is that it is not possible to align the memory loads on natural alignment boundaries, i.e., using memory addresses divisible by the length of the register in bytes.
However, most modern processor designs
can handle unaligned SIMD load and store operations at little to no performance penalty.  

Another side effect of our approach is that loop-carried dependencies are eliminated entirely, making it easier for  the processor to overlap multiple iterations. In theory, given enough execution units, a processor could execute several iterations simultaneously. 

The third choice is based on the expectation that almost all text processed by our procedure is correctly sequenced.  Therefore, it is advantageous to skip the expensive step of determining the incorrect surrogates and fixing them up unless a quick check shows a need to do so.  As a bonus, for in-place operation, writing to the buffer can be avoided entirely for valid UTF-16.

\begin{figure}
    \centering
    \begin{tabular}{l}
    \lstinputlisting[style=customcpp]{code/avx512\_utf16.c}
    \end{tabular}
    \caption{AVX-512 function to replace invalid UTF-16 surrogates with the replacement character within a 64-byte block}
    \label{fig:avx512simd}
\end{figure}

\subsection{AVX-512 (x64) implementation}

Intel and AMD 64-bit processors (x86-64), which dominate the market for Windows PCs and servers, support a range of SIMD instruction set extensions, falling into the four families of MMX (64-bit vectors, of historical interest only), SSE (128-bit vectors), AVX (256-bit vectors), and AVX-512 (512-bit vectors).
Within each family, there are multiple levels of support, with each level including all previous levels as well as all previous families.
All x64 processors support at least SSE2, though it is generally advantageous to make use of the widest SIMD extension available to maximize the amount of data processed per operation.
SSE and its extensions (SSE2, SSE3, SSSE3, SSE4.1, SSE4.2) use 128-bit XMM registers, capable of handling 16~bytes of data (e.\,g., four 32-bit floats or sixteen 8-bit integers). AVX and AVX2 introduce 256-bit YMM registers, doubling the capacity to 32~bytes (e.g., eight 32-bit floats). AVX-512 further expands to 512-bit ZMM registers, processing 64~bytes (e.\,g., sixteen 32-bit floats) and introduces new features like mask registers.

All these instruction sets provide fast comparison instructions capable of comparing chunks of 16-byte (SSE), 32-byte (AVX/AVX2), or 64-byte (AVX-512) data at the 16-bit granularity, making them well suited for working with UTF-16 data.
In practice, the variety of instruction sets available on x64 processors often requires runtime detection of supported instruction sets.  For example, the first time the processing is initiated, we might check for the supported CPU features and pick one out of several precompiled functions.

For instruction sets prior to AVX-512, comparison instructions generate results in SIMD registers as either all-ones (\texttt{0xFF\ldots{}FF}) or all-zeros (\texttt{0x00\ldots{}00}) per element, indicating true or false outcomes. These results can be efficiently mapped to a general-purpose register acting as a bitset, where each bit corresponds to a comparison result from the SIMD register. For example, with the SSE2 \texttt{pcmpeqw} instruction, comparing two 16-byte vectors produces a 128-bit XMM register with 16~bytes, where each 16-bit subword is either \texttt{0xFFFF} or \texttt{0x0000} depending on whether the comparison was true. Using the SSE2 \texttt{pmovmskb} instruction, these bytes are converted into a 16-bit integer, where each bit represents one byte’s comparison result. This bitset can then be processed in scalar code. Example: comparing two vectors of 16 bytes might yield an XMM register with bytes [\texttt{0xFF}, \texttt{0xFF}, \texttt{0x00}, \texttt{0x00}, \texttt{0xFF}, \texttt{0xFF}, \ldots{}], and we can map this to a 16-bit value like \texttt{0b110011}\ldots{}, where 1 indicates equal bytes. The process is much the same for AVX2: e.g., the AVX2 \texttt{vpmovmskb} instructions act like the SSE2 \texttt{pmovmskb} instruction but produce a 32-bit register value instead of a 16-bit register value.

AVX-512 introduces dedicated mask registers (k0–k7), which replace the need for movemask operations (\texttt{pmovmskb} and \texttt{vpmovmskb}) in many cases. These 8-bit to 64-bit mask registers directly store comparison results, with each bit corresponding to an element in the 512-bit ZMM register. For instance, a comparison like the \texttt{vpcmpw} instructions on two 512-bit vectors (thirty-two 16-bit integers) produces a 32-bit mask in an AVX-512 mask register, where each bit indicates equality for one integer. These masks can be used directly in subsequent AVX-512 instructions for conditional operations or merged with general-purpose registers for scalar processing. This eliminates the overhead of movemask instructions, with the caveat that operations on the mask registers are limited (e.g., there are no trailing or counting zero instructions).

\paragraph{Architecture-specific adjustments}
With SSE2 and AVX2, the generic procedure can be used without changes.
It is also applicable to AVX-512, but benefits from using a specific \texttt{utf16fix\_block} function with 
some architecture-specific changes,
see Fig.~\ref{fig:avx512simd}.
The AVX-512 implementation of the \texttt{utf16fix_block} function leverages 512-bit ZMM registers to process 
thirty-two 
UTF-16 code units (64~bytes) in parallel, utilizing AVX-512's mask registers for efficient conditional 
operations.
Other than being cast in architecture-specific intrinsics, these are the differences in 
\texttt{utf16fix\_block} procedure compared to our generic algorithm:

\begin{itemize}
\item instead of using vectors of \texttt{FFFF} or \texttt{0000} for boolean results, we use mask registers,
\item the various blend operations are realised by mask-driven blend instructions and masked stores instead of bit-operations,
\item instead of fixing up $\emph{out}[-1]$ by extracting \emph{lb} from \emph{lb\_illseq} into a scalar, we perform a 64~byte store into $\emph{out}-1$ masked with just the first bit of \emph{lb\_illseq}, guaranteeing that it stores either one code unit or none and in particular does not overlap the subsequent store to \emph{out}.
\end{itemize}
For simplicity, we omit the in-place operation.
In production, the case of $\emph{in}=\emph{out}$ is sped up by replacing the blend step in the correction path 
with a straight masked store to only the illegally sequenced surrogates and by leaving out the store in the 
fast path entirely.

Table~\ref{tab:simd_instructions} summarizes the SIMD instructions and their corresponding intrinsics used for UTF-16 processing, as detailed in the provided code and discussion. The table lists each instruction, its associated intrinsic, the instruction set (SSE2, AVX2, or AVX-512), and a brief description of its functionality. Instructions like \texttt{pcmpeqw} and \texttt{pmovmskb} from SSE2 handle 128-bit vectors, while AVX2's \texttt{vpmovmskb} extends to 256-bit vectors. AVX-512 instructions, such as \texttt{vpcmpw} and \texttt{vpblendmw}, leverage 512-bit ZMM registers and mask registers for efficient 32-element processing, with intrinsics like \texttt{\_mm512_cmpeq_epi16_mask} and \texttt{\_kxor_mask32} enabling  surrogate pair validation.

\begin{table}
\centering
\caption{x64 SIMD instructions and intrinsics for UTF-16 processing}
\label{tab:simd_instructions}\footnotesize
\begin{tabular}{l l l p{6cm}}
\toprule
Instruction & Intrinsic & Instruction Set & Description \\
\midrule
pcmpeqw & \texttt{_mm_cmpeq_pi16} & SSE2 & Compares 16-bit integers in two 128-bit vectors, setting each 16-bit element to \texttt{0xFFFF} (true) or \texttt{0x0000} (false) if equal. \\
pmovmskb & \texttt{_mm_movemask_pi8} & SSE2 & Extracts the most significant bit of each byte in a 128-bit vector, producing a 16-bit integer bitmask. \\
vpmovmskb & \texttt{_mm256_movemask_epi8} & AVX2 & Extracts the most significant bit of each byte in a 256-bit vector, producing a 32-bit integer bitmask. \\
vpcmpw & \texttt{_mm512_cmpeq_epi16_mask} & AVX-512 & Compares 16-bit integers in two 512-bit vectors, producing a 32-bit mask where each bit indicates equality. \\
vmovdqu & \texttt{_mm512_loadu_si512} & AVX-512 & Loads 512 bits of unaligned data into a ZMM register. \\
vpandd & \texttt{_mm512_and_epi32} & AVX-512 & Performs a bitwise AND on two 512-bit vectors, treating data as 32-bit integers. \\
kxord & \texttt{_kxor_mask32} & AVX-512 & Performs a bitwise XOR on two 32-bit mask registers. \\
ktestzq & \texttt{_ktestz_mask32_u8} & AVX-512 & Tests if all bits in a 32-bit mask are zero, returning true if so. \\
kandnw & \texttt{_kandn_mask32} & AVX-512 & Performs a bitwise AND NOT on two 32-bit mask registers (inverts first mask, then ANDs). \\
korw & \texttt{_kor_mask32} & AVX-512 & Performs a bitwise OR on two 32-bit mask registers. \\
kshiftriw & \texttt{_kshiftri_mask32} & AVX-512 & Shifts a 32-bit mask right by one bit, inserting a zero at the most significant bit. \\
kmovw & \texttt{_cvtu32_mask32} & AVX-512 & Converts an unsigned 32-bit integer to a 32-bit mask register. \\
vmovdqu16 & \texttt{_mm512_mask_storeu_epi16} & AVX-512 & Stores 16-bit integers from a 512-bit vector to unaligned memory, using a 32-bit mask to select elements. \\
vmovdqu & \texttt{_mm512_storeu_si512} & AVX-512 & Stores a 512-bit vector to unaligned memory. \\
vpblendmw & \texttt{_mm512_mask_blend_epi16} & AVX-512 & Blends 16-bit integers from two 512-bit vectors based on a 32-bit mask. \\
vmovdqu32 & \texttt{_mm512_storeu_epi32} & AVX-512 & Stores 32-bit integers from a 512-bit vector to unaligned memory. \\
vpbroadcastw & \texttt{_mm512_set1_epi16} & AVX-512 & Broadcasts a single 16-bit integer to all elements of a 512-bit vector. \\
vpbroadcastd & \texttt{_mm512_set1_epi32} & AVX-512 & Broadcasts a single 32-bit integer to all elements of a 512-bit vector. \\
\bottomrule
\end{tabular}
\end{table}

\begin{figure}[p]
    \centering
    \begin{tabular}{l}
    \lstinputlisting[style=customcpp]{code/neon\_utf16.c}
    \end{tabular}
    \caption{ARM NEON implementation for UTF-16 correction}
    \label{fig:neon}
\end{figure}
\afterpage{\clearpage}

\subsection{NEON (Aarch64) implementation}
Aarch64 processors are  common on mobile devices like smartphones and tablets as well as appliances such as Smart TVs and videogame consoles.  Recently, they are becoming more and more common in the server market and, with Apple switching to Aarch64, also on desktops and laptops.
Two SIMD instruction set extensions are specified for Aarch64.
The NEON extension taken over and extended from the 32~bit ARM architecture provides 128-bit registers capable of processing eight 16-bit words simultaneously.
Its feature set is comparable to that of SSSE3 with a full set of integer and floating point operations as well as loads, stores, and shuffles.
NEON support is mandatory on all Aarch64 processors with few exceptions and thus the most common target for SIMD-acceleration on Aarch64.
The newer SVE/SVE2/SVE2.1 family of instruction set extensions builds on NEON and extends it with variable-length vector registers, predicate mask and many other features, rendering it similar to AVX-512.
Unfortunately, adoption of SVE has been slow, with almost all OEMs opting to ship CPUs that have no support.
We have therefore decided to ignore SVE and use NEON.

In contrast to x64, hardware supporting NEON has  diverse execution characteristics.
A low-power Aarch64 CPU might only be able to execute one NEON instruction every three cycles.
High-power CPUs, on the other hand, have multiple execution units, often more than x64 processors.
The resulting higher throughput compensates for the shorter vector length.
For example, the Firestorm Cores of Apple M1 processors can execute four of most NEON instructions
per cycle~\cite{applesilicon}.

\paragraph{Architecture-specific adjustments}
Fig.~\ref{fig:neon} shows our NEON implementation.
While it is possible to use the generic algorithm (Fig.~\ref{fig:genericsimd}) on NEON, there are two important changes to be made that significantly improve performance.
First, it should be noted that NEON does not have an equivalent to SSE's and AVX' mask-moving instructions such as `pmovmskb`.
It also lacks a dedicated instruction to check that a register is non-zero.
We found an efficient workaround to check that a 128-bit value is non-zero by computing a vertical maximum 
(maximum over all elements) using an instruction such as \texttt{vmaxvq_u32} and then moving the result to a 
general-purpose register, which holds zero if and only if all elements of the vector were zero.
The performance of these instructions varies depending on the chosen hardware, but both instructions can have 
several cycles of latency. Therefore, this step can become a bottleneck.
We can implement a \texttt{utf16fix\_block} function which works similarly to the x64 function, but the expected performance is not ideal.

We amortize this cost by processing 4 blocks in parallel and only then checking if any of the blocks require correction.
Only if this is the case, do we branch to fix up the illegal surrogates.
As we expect most inputs to not require correction (most UTF-16 inputs are valid), we can alleviate the problem by processing the data in larger blocks, spanning 64~code points.
As a side effect, the instruction-level parallelism is increased, benefiting microarchitectures with many execution ports like the Apple M1.

The other important change is to make use of the NEON-exclusive deinterleaving load instruction \texttt{vld2q_u8}.
The LD2 instruction loads two vectors' worth of bytes from memory, writing the even-numbered bytes to one register and the odd-numbered bytes to another.\footnote{A companion instruction \texttt{vst2q_u8} can undo this transformatin on store, but is not needed here.}
With \texttt{vld2q_u8}, it is advantageous to treat UTF-16 code units as pairs of bytes.
As the more significant of the two bytes in a UTF-16 code unit suffices to tell if the code unit is a high surrogate, low surrogate, or neither, we use \texttt{vld2q_u8} to load 16~code units into a pair of vectors, and then discard the vector of the less significant bytes as its contents are not required to carry out the algorithm.
We carry on with just the high-byte vector, having now reduced the element size from 16 to 8~bits, doubling the number of code units processed per step.
As a side effect, having discarded the low-byte vector, we cannot fix up the vectors as in Fig.~\ref{fig:genericsimd}, and use a different approach instead.

Like with the other implementations, the core function, \texttt{to_well_formed}, orchestrates the processing of 
a UTF-16 string of length~$n$.
It begins by handling short inputs ($n<17$) with a scalar fallback.
For longer inputs, it ensures the first output element is not an invalid low surrogate by replacing it with \texttt{U+FFFD} if necessary.
The function then processes the string in blocks of 64~code points using \texttt{utf16fix_block64} when possible, falling back to blocks of 16~code points with \texttt{utf16fix_block} for smaller remaining segments.
A final 16-code-point block is processed to cover the string’s end, and the last element is checked to ensure it is not an invalid high surrogate, replacing it with~\texttt{U+FFFD} if needed. 

The \texttt{utf16fix_block64} function processes sixty-four UTF-16 code units (128-bytes) by dividing the block into four 16-code-unit segments, each handled by \texttt{get_mismatch_copy}.
This helper function loads two 128-bit vectors: \texttt{lb} (lookback, from $\emph{in} - 1$) and \texttt{block} (from \emph{in}), using \texttt{vld2q_u8} to deinterleave bytes into high and low parts.
The function then masks the high bytes with~\texttt{0xFC} to focus on the top 6 bits, identifying high surrogates (\texttt{0xD8}) and low surrogates~(\texttt{0xDC}) via a comparison instruction \texttt{vceqq_u8}.
An exclusive-or operation (\texttt{veorq_u8}) detects illegal sequences (e.\,g., a high surrogate not followed by a low surrogate).
The function copies the input block to the output using \texttt{vst2q_u8} and returns the illegal sequence mask. In \texttt{utf16fix_block64}, four such masks (\emph{illse0} to \emph{illse3}) are combined using \texttt{vorrq_u8} and checked for non-zero status with \texttt{veq_non_zero}, which uses \texttt{vmaxvq_u32} to compute the maximum across a 128-bit vector, efficiently detecting any invalid sequences.

If invalid sequences are detected, \texttt{get_mask} converts the four 128-bit masks into a 64-bit bitset, 
where each bit corresponds to a code unit’s validity. It uses a predefined \emph{bit\_mask} vector to extract 
specific bits, accumulating results with \texttt{vpaddq_u8} to produce a single 64-bit value via \texttt{vgetq_lane_u64}. This bitset is processed scalarly, using \texttt{stdc_trailing_zeros_ull} to find invalid positions, determining whether each is a high or low surrogate, and replacing the appropriate code unit with~\texttt{U+FFFD}. The function returns false to indicate corrections were made, or true if the block was valid and copied unchanged. The \texttt{utf16fix_block} function is omitted for simplicity; it works similarly to the generic implementation.
The main difference being that it uses interleaved loads.

Table~\ref{tab:neon_instructions} lists the ARM NEON instructions and their corresponding intrinsics used in the UTF-16 processing code. These instructions operate on 128-bit NEON registers, enabling parallel processing of eight 16-bit words. Each entry includes the instruction, its intrinsic, and a concise description of its role in validating and correcting UTF-16 surrogate pairs.
\begin{table}[tbh]
\centering
\caption{ARM NEON instructions and intrinsics for UTF-16 processing}
\label{tab:neon_instructions}
\footnotesize
\begin{tabular}{r l p{6cm}}
\toprule
Instruction & Intrinsic & Description \\
\midrule
\texttt{LD2} & \texttt{vld2q\_u8} & Loads 256 bits of unaligned data from memory, deinterleaving into two 
128-bit vectors of 8-bit integers. \\
\texttt{ST2} & \texttt{vst2q\_u8} & Stores two 128-bit vectors of 8-bit integers to unaligned memory, 
interleaving the data. \\
\texttt{AND} & \texttt{vandq\_u8} & Performs a bitwise AND on two 128-bit vectors of 8-bit integers. \\
\texttt{CMEQ} & \texttt{vceqq\_u8} & Compares two 128-bit vectors of 8-bit integers for equality, producing a 
128-bit vector with \texttt{0xFF} for true and \texttt{0x00} for false per element. \\
\texttt{EOR} & \texttt{veorq\_u8} & Performs a bitwise XOR on two 128-bit vectors of 8-bit integers. \\
\texttt{DUP} & \texttt{vdupq\_n\_u8} & Broadcasts a single 8-bit integer to all elements of a 128-bit vector. 
\\
\texttt{UMAXV} & \texttt{vmaxvq\_u32} & Computes the maximum value across all 32-bit elements in a 128-bit 
vector, returning a 32-bit scalar. \\
\texttt{ADDP} & \texttt{vpaddq\_u8} & Pairwise adds 8-bit integers from two 128-bit vectors, producing a 
128-bit vector of 8-bit sums. \\
\texttt{FMOV} & \texttt{vgetq\_lane\_u64} & Extracts a 64-bit lane from a 128-bit vector, interpreting the 
vector as two 64-bit integers. \\
n/a & \texttt{vreinterpretq\_u64\_u8} & Reinterprets a 128-bit vector of 8-bit integers as a 128-bit vector of 
64-bit integers without changing the data. \\
n/a & \texttt{vreinterpretq\_u32\_u8} & Reinterprets a 128-bit vector of 8-bit integers as a 128-bit vector of 
32-bit integers without changing the data. \\
\bottomrule
\end{tabular}
\end{table}

\section{Experiments}
To evaluate the performance of UTF-16 validation and correction algorithms, we developed a  benchmarking tool in C++ (\texttt{benchmark_to_well_formed_utf16.cpp}). It is part of the \texttt{simdutf} library.\footnote{\url{https://github.com/simdutf/simdutf}, commit hash~\texttt{dae7a10}}. We build the benchmarking software and the C++ library in release mode (\texttt{-O3 -NDEBUG}).
On x64 processors, the simdutf library uses specialized kernels optimized for different Intel processor microarchitectures, including icelake, haswell, and westmere. The icelake kernel is tailored for Intel’s Ice Lake processors (AVX-512 with VBMI2). The haswell kernel is optimized for the  Haswell microarchitecture with AVX2 support. The westmere kernel targets older Westmere processors (SSE4.2).

We present the systems we use for benchmarking in Table~\ref{tab:test-cpus}.
This table details the key specifications of each system, including processor type, clock frequency, microarchitecture, memory configuration, and compiler version. The selected systems represent a mix of modern high-performance architectures, allowing for a comprehensive evaluation of performance across different workloads and computational environments.

 \begin{table}
   \caption{Systems used for benchmarking}%
   \label{tab:test-cpus}
   \centering
   
  \begin{tabular}{rll}
  \toprule
  Processor&Apple M4&Intel Xeon Gold 6338\\
  \midrule
  Frequency&\SIrange{4.4}{4.5}{\GHz}&\SI{3.2}{\GHz}\\
  Microarchitecture&M4 (aarch64, 2024)&Ice~Lake (x64, 2019)\\
  Memory&LPDDR5X (7500\,MT/s)&DDR4 (3200\,MT/s)\\
  Compiler&Apple/LLVM 17&GCC 12\\
  \bottomrule
  \end{tabular}
 \end{table}

We generate random UTF-16 strings with controlled characteristics. Specifically, we configure the input strings with a specified percentage of valid surrogate pairs (code units \texttt{U+D800} to \texttt{U+DBFF} followed by \texttt{U+DC00} to \texttt{U+DFFF}) and mismatched surrogates (isolated high or low surrogates). Specifically, if we set the percentage of surrogate pairs to \SI{0.1}{\percent}, then \SI{0.1}{\percent} of all characters involve surrogate pairs. If we set the percentage of mismatched surrogates to \SI{0.1}{\percent}, then \SI{0.1}{\percent} of the code units outside the valid surrogate pairs are randomly chosen (\SI{50}{\percent}) high or low surrogates. It is possible, but unlikely, that some isolated surrogates might form valid pairs (i.e., for \SI{0.1}{\percent} the probability is \SI{0.01}{\percent}).

The benchmark measures the throughput (in \unit{\giga\byte\per\second}) and hardware performance counters, such as instructions per byte and cycles per byte, for different implementations, including a baseline from the V8 JavaScript engine and optimized versions using the simdutf library. 
The V8 code was replaced by the simdutf library in recent versions.
The experiments are conducted with input sizes of up to \num{1000000}~code units.

We conducted experiments with two distinct configurations to analyze the algorithms' behavior under different conditions. In the first configuration, we set the input size to \num{1000000}~code units, with \SI{0.1}{\percent} of the code units forming valid surrogate pairs and \SI{0}{\percent} as mismatched surrogates. This setup represents a scenario with minimal supplementary plane characters, focusing on basic UTF-16 characters. In the second configuration, we maintained the same input size but adjusted the parameters to include \SI{0.1}{\percent} valid surrogate pairs and \SI{0.1}{\percent} mismatched surrogates, introducing a small proportion of invalid UTF-16 sequences. These configurations were chosen to evaluate the performance of the algorithms in both nearly valid and slightly erroneous UTF-16 inputs, providing insights into their efficiency and error-handling capabilities.

\begin{figure}[H]
    \centering
    \includegraphics[width=0.5\linewidth]{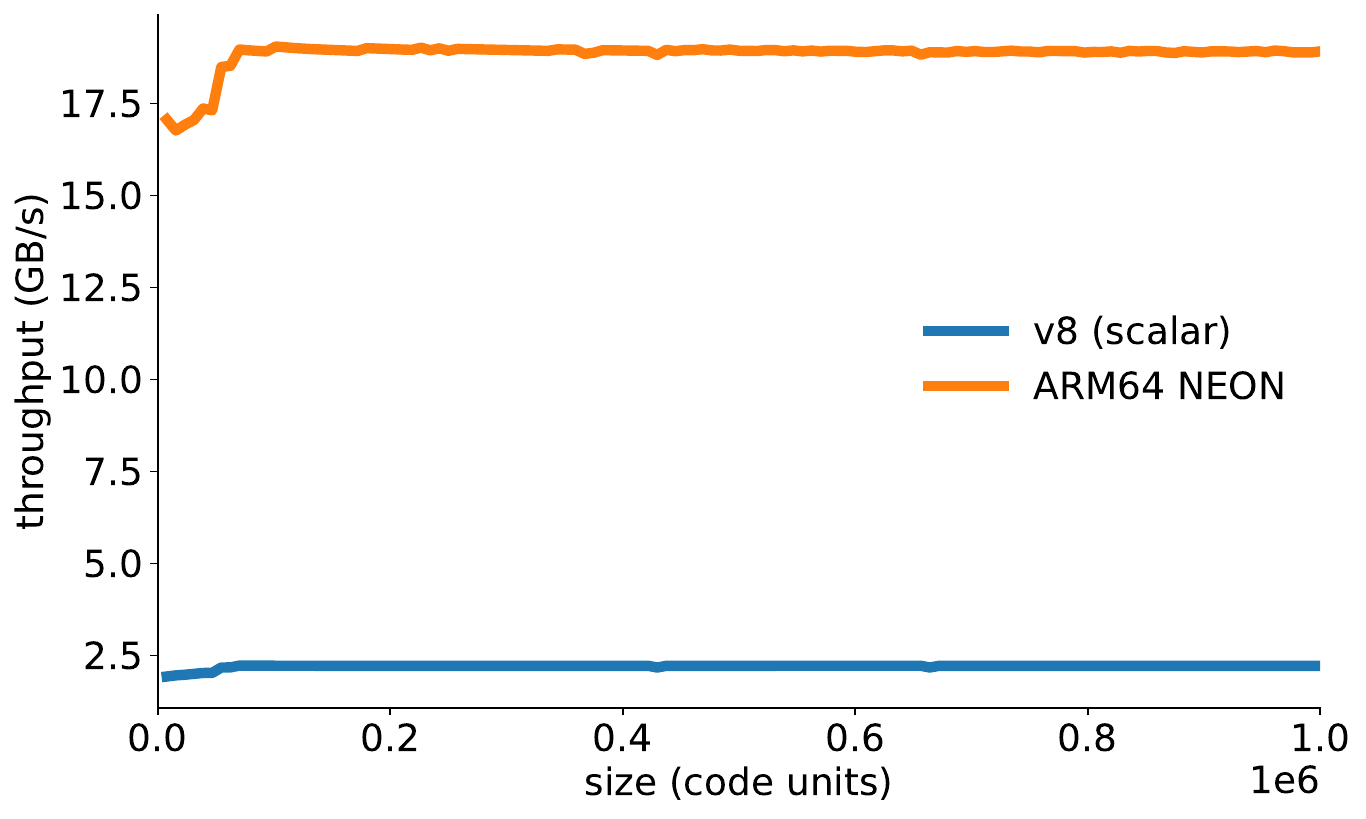}
    \caption{Throughput on the Apple system for 0 to a million code units}
    \label{fig:throughput}
\end{figure}

In addition to the fixed-size experiments, we performed benchmarks across a range of input sizes to assess scalability. The framework divides the maximum input size (\num{1000000}~code units) into 128~nearly equal chunks, testing each chunk independently. This approach ensures a fine-grained analysis of performance trends as the input size increases. For each chunk, the benchmark generates a random UTF-16 string with the specified surrogate pair and mismatched surrogate percentages (\SI{0.1}{\percent} and \SI{0}{\percent} or \SI{0.1}{\percent} and \SI{0.1}{\percent}, depending on the configuration). The performance metrics, including throughput and error margins, are collected for each implementation, allowing us to compare their behavior across different input scales and identify any size-dependent bottlenecks.
We present this result in Fig.~\ref{fig:throughput} for the Apple platform. We see that the scalar version (V8) and the simdutf version (ARM64) have consistent speeds throughout the range, although they are both slightly slower on a per \si{\giga\byte\per\second} basis for tiny strings. The simdutf function is nearly 9~times faster than the scalar function.

Each implementation is tested over 100~iterations, with the input data processed multiple times to achieve measurable execution times (at \SI{1}{\milli\second}). The software uses an event collector to capture both elapsed time and hardware counters, such as CPU cycles and instructions, when available. The throughput is calculated as the input size divided by the best execution time (in nanoseconds), reported in \si{\giga\byte\per\second}. Additionally, we compute the error margin as the percentage difference between the average and best execution times, providing a measure of performance stability. We find a low error margin (about 1\%).

The performance results for the Apple and Intel systems are summarized in Table~\ref{tab:apple} and Table~\ref{tab:intel}, respectively. Table~\ref{tab:apple} shows the throughput (\si{\giga\byte\per\second}), instructions per byte, and instructions per cycle for the V8 baseline and the optimized arm64 implementation on the Apple M4 system, under both \SI{0}{\percent} and \SI{0.1}{\percent} mismatched surrogate conditions. The arm64 implementation significantly outperforms the V8 baseline, achieving up to \SI{18.9}{\giga\byte\per\second} with an error rate of \SI{0}{\percent} compared to \SI{2.2}{\giga\byte\per\second} for V8, demonstrating the effectiveness of SIMD optimizations. Table~\ref{tab:intel} presents similar metrics for the Intel Xeon Gold 6338 system, comparing the V8 baseline with optimized implementations for Ice Lake, Haswell, and Westmere microarchitectures. The Ice Lake implementation achieves the highest throughput at \SI{7.5}{\giga\byte\per\second} with \SI{0}{\percent} errors, while the V8 baseline lags at \SI{1.2}{\giga\byte\per\second}. Across both systems, the optimized implementations exhibit lower instruction counts per byte and higher efficiency, particularly in error-free scenarios, highlighting the benefits of architecture-specific SIMD optimizations. The performance comparison between icelake and haswell kernels, in Table~\ref{tab:intel}, reveals distinct differences in throughput and efficiency on the Intel Xeon Gold 6338 system. The icelake kernel achieves a throughput of \SI{7.5}{\giga\byte\per\second} at \SI{0}{\percent} error rate, slightly below Haswell’s \SI{7.8}{\giga\byte\per\second} under the same condition. However, the icelake kernel demonstrates superior efficiency, with a lower instruction count per byte (0.4 versus haswell’s 0.8) and significantly lower instructions per cycle (1.0 compared to haswell’s 1.8).

\begin{table}
\centering
\caption{Performance metrics for Apple systems at size \num{1000000}\label{tab:apple}}
\begin{tabular}{lcccc}
\toprule
& \multicolumn{2}{c}{V8} & \multicolumn{2}{c}{arm64} \\
\cmidrule(lr){2-3} \cmidrule(lr){4-5}
Error & \SI{0}{\percent} & \SI{0.1}{\percent} & \SI{0}{\percent} & \SI{0.1}{\percent} \\
\midrule
\si{\giga\byte\per\second} & 2.2 & 2.2 & 18.9 & 16.3 \\
ins/byte & 12.0 & 12.0 & 0.9 & 0.9 \\
ins/cycle & 5.9 & 5.8 & 3.7 & 3.3 \\
\bottomrule
\end{tabular}
\end{table}

\begin{table}
\centering
\caption{Performance metrics for Intel systems at size \num{1000000}\label{tab:intel}}
\begin{tabular}{lcccccccc}
\toprule
& \multicolumn{2}{c}{V8} & \multicolumn{2}{c}{icelake} & \multicolumn{2}{c}{haswell} & \multicolumn{2}{c}{westmere} \\
\cmidrule(lr){2-3} \cmidrule(lr){4-5} \cmidrule(lr){6-7} \cmidrule(lr){8-9}
Error & \SI{0}{\percent} & \SI{0.1}{\percent} & \SI{0}{\percent} & \SI{0.1}{\percent} & \SI{0}{\percent} & \SI{0.1}{\percent} & \SI{0}{\percent} & \SI{0.1}{\percent} \\
\midrule
\si{\giga\byte\per\second} & 1.2 & 1.2 & 7.5 & 7.4 & 7.8 & 7.6 & 5.8 & 5.6 \\
ins/byte & 13.0 & 13.0 & 0.4 & 0.4 & 0.8 & 0.8 & 2.0 & 2.0 \\
ins/cycle & 5.0 & 5.0 & 1.0 & 1.0 & 1.8 & 1.8 & 3.6 & 3.5 \\
\bottomrule
\end{tabular}
\end{table}

\subsection{JavaScript Performance}

Our C++ functions have been integrated in the popular Node.js JavaScript runtime engine as of version 25. Prior versions relied a conventional implementation of the toWellFormed function. Using a 1024-character random string made of 5\% of low surrogate, 5\% high surrogate and otherwise filled with ASCII characters, we get the results in Table~\ref{tab:js}.\footnote{See \url{https://github.com/lemire/browserwellformed} for our JavaScript benchmark code.} Under both systems, we find a five-fold performance gain, consistent with our C++ benchmarks.

\begin{table}
\centering
\caption{Performance in Node.js of the toWellFormed function (random 1024-character string)\label{tab:js}}
\begin{tabular}{lcc}
\toprule
Node.js version & Apple M4 & Intel Xeon Goald 6338 \\
\midrule
24.13.0 & \SI{2.9}{\gibi\byte\per\second} &\SI{2.0}{\gibi\byte\per\second} \\
25.5.0 &\SI{16}{\gibi\byte\per\second} &\SI{11}{\gibi\byte\per\second} \\
\bottomrule
\end{tabular}
\end{table}

\section{Conclusion}
 By leveraging SIMD instructions, our proposed algorithm achieves significant performance improvements over traditional scalar methods, with up to eightfold speedups on ARM NEON and x64 SSE architectures. Experimental results on Apple M4 and Intel Xeon systems confirm the scalability and effectiveness of our approach, with throughputs reaching \SI{18.9}{\giga\byte\per\second} on Apple~M4 processor and \SI{7.5}{\giga\byte\per\second} on an Intel~Ice~Lake processor, alongside reduced instruction counts. 

\section{Funding}
This research was funded by the Natural Sciences and Engineering Research Council of Canada under grant number RGPIN-2024-03787 and by the NHR~e.\,V. Graduate School Programme.
\bibliography{utf16_fix}
\end{document}